\documentclass[twocolumn,showpacs,preprintnumbers]{revtex4}
\usepackage{graphicx}
\usepackage{dcolumn}
\usepackage{bm}
\usepackage{amsmath}

\begin{document}

\title{Laser-like X-ray Sources Based on Optical Reflection from Relativistic Electron Mirror}
\author{H.-C. Wu$^{1,}$\footnote{hcwu@lanl.gov}, J. Meyer-ter-Vehn$^{2}$, J. Fern\'{a}ndez$^{1}$, and B.M.
Hegelich$^{1,3}$}
\affiliation{$^{1}$Los Alamos National Laboratory, Los Alamos, New Mexico 87545, USA\\
$^{2}$Max-Planck-Institut f\"{u}r Quantenoptik, D-85748 Garching, Germany\\
$^{3}$Department f\"{u}r Physik, Ludwig-Maximilians-Universit\"{a}t M\"{u}%
nchen, D-85748 Garching, Germany}
\date{\today }

\begin{abstract}
A novel scheme is proposed to generate uniform relativistic electron layers
for coherent Thomson backscattering. A few-cycle laser pulse is used to
produce the electron layer from an ultra-thin solid foil. The key element of
the new scheme is an additional foil that reflects the drive laser pulse,
but lets the electrons pass almost unperturbed. It is shown by analytic
theory and by 2D-PIC simulation that the electrons, after interacting with
both drive and reflected laser pulse, form a very uniform flyer freely
cruising with high relativistic $\gamma$-factor exactly in drive laser
direction (no transverse momentum). It backscatters probe light with a full
Doppler shift factor of $4\gamma^2$. The reflectivity and its decay due to
layer expansion is discussed.
\end{abstract}

\maketitle

\pagebreak

High-quality X-ray sources are requested in many fields of science.
Presently, large free-electron lasers (FEL) \cite{Ackermann2007} represent
powerful coherent VUV and X-ray sources, which open a new era of intense VUV
or X-ray interaction with matter and provide unprecedented opportunities for
research in condensed matter physics, high-energy-density physics \cite%
{Lee2003} and single biomolecular imaging \cite{Neutze2000}. High laser
harmonics from gas targets \cite{Seres2005} and relativistic laser plasma
interaction \cite{Dromey2007} are also very useful and promising coherent
X-ray sources. Such harmonic sources typically produce trains of sharp
spikes separated by the time period of the driving laser pulse.

Bright and coherent X-ray sources can also be obtained by coherent Thomson
scattering (CTS) from dense electron layers flying with relativistic factor $%
\gamma _{x}=1/\sqrt{1-\beta _{x}^{2}}$. Here $\beta _{x}=v_{x}/c$ is the
velocity of the plane flyer in normal direction. Counter-propagating probe
light is then mirrored and frequency-upshifted by the relativistic Doppler
factor, which is $(1+\beta _{x})/(1-\beta _{x})\approx 4\gamma _{x}^{2}$ for
$\gamma _{x}\gg 1$ \cite{Einstein1905}. In this paper we refer to these
electron layers as relativistic electron mirrors (REM) or simply flyers. One
way to produce them is to drive cold non-linear plasma waves to the point of
wave breaking. Their density profile then develops diverging spikes that may
move with high $\gamma _{x}$-factors \cite{Bulanov2003}. Recent experiments
have demonstrated $\gamma _{x}\approx 5$ by identifying the mirrored light
\cite{Pirozhkov2007}. This method is limited by the phase velocity of the
plasma wave requiring low plasma density for high $\gamma _{x}$-factors of
the wave. Higher densities can be achieved by accelerating thin solid foils.
Corresponding simulations were reported recently \cite{Esirkepov2009},
driving a 250 nm thick foil with a laser intensity of about $10^{23}$W/cm$%
^{2}$. Only small $\gamma _{x}$ values are obtained in this case, because
the complete foil including ions is accelerated.

In this letter, a different regime is considered requiring much lower laser
intensities and foils thin enough for the laser pulse to push out all
electrons from the foil. In this case, only electrons are accelerated and
can gain high $\gamma $ values, while the heavy ions are left behind
unmoved. For this to happen, the normalized laser field $a_{0}=eE_{L,0}/mc%
\omega _{L}$ has to be much larger than the normalized field arising from
charge separation, $E_{x,0}=(n_{e}/n_{c})k_{L}d $. Here $\omega _{L}=ck_{L}$
is the circular frequency of the driving laser pulse, $n_{e}$ is the
electron density and $d$ the thickness of the foil initially, while $%
n_{c}=\epsilon _{0}m_{e}\omega _{L}^{2}/e^{2}$ is the critical density. This
regime has been described by Kulagin in a number of papers (see e.g.\cite%
{Kulagin}), and coherent Thomson scattering (CTS) from the emerging
relativistic electron layers has been studied in Refs. \cite{CTS1,CTS2}. Some
typical results are shown in Fig. 1. The scheme requires extremely thin
foils of a few nanometer and laser pulses with a very high contrast ratio in
order not to destroy the target before the main pulse arrives. These foils
are much thinner than the skin depth and are therefore transparent, even
though the electron density is overcritical. The leading edge of the laser
pulse ionizes the foil and takes the electrons along as a thin sheath (see
Fig. 1(a) and Fig. 2). Their motion is well described in a single-electron
picture \cite{MtV2001,Wen2009}; transverse and longitudinal momenta follow
approximately the local laser vector potential $a(x,t)$ according to $%
p_{\perp }=a$, $p_{x}=\gamma -1=a^{2}/2$, where $\gamma $ is now the full
relativistic factor.

\begin{figure}[tbp]
\resizebox{1\columnwidth}{!}{\includegraphics{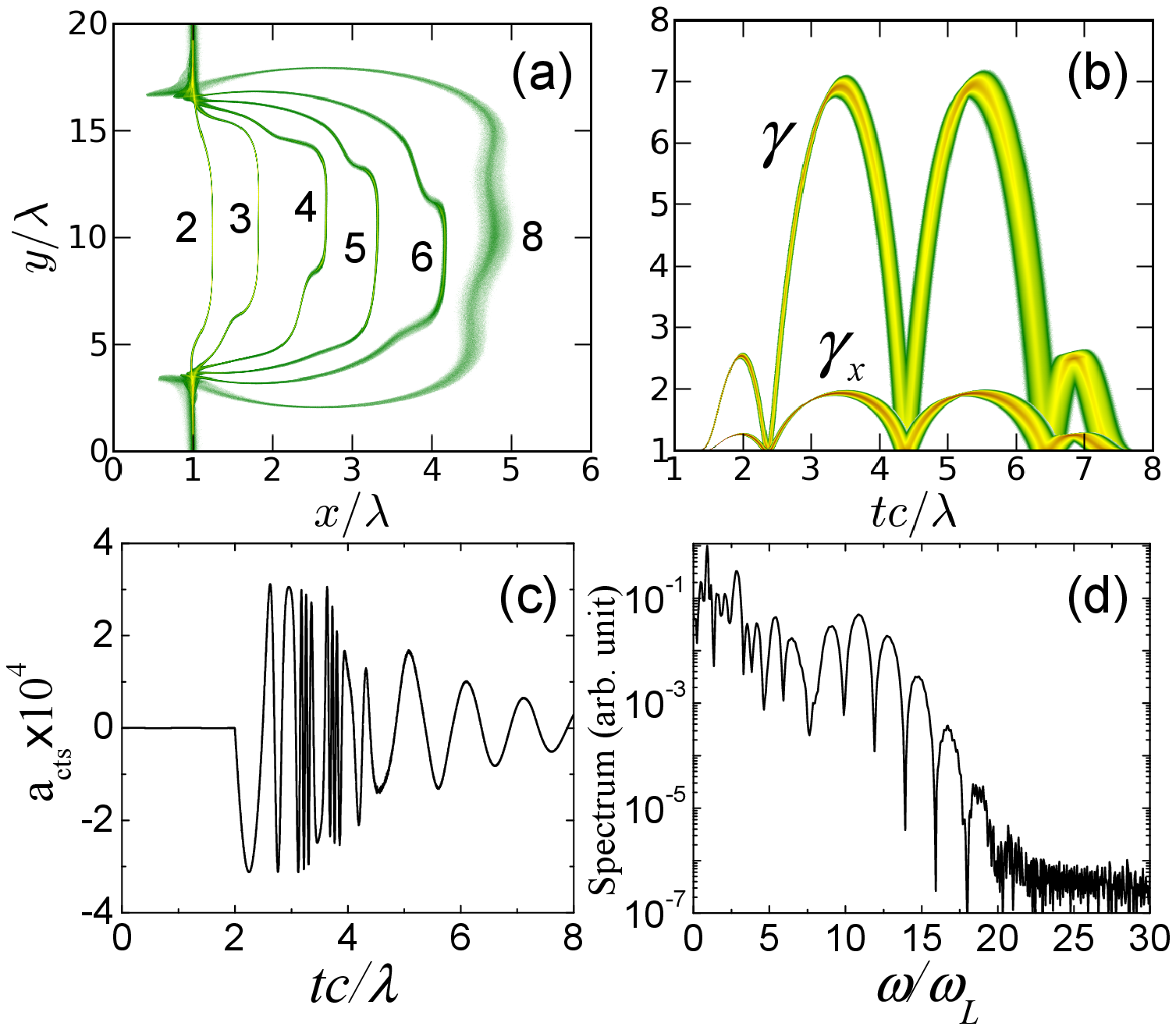}}
\caption{(color). Laser-driven electron layer blown out from a thin foil at $%
x/\protect\lambda=1$ by a drive laser pulse incident from the left in $(+x)$%
-direction and reflected light from a probe pulse incident from the right in
$(-x)$-direction. (a) Simulated electron density in (x,y) plane given at
different times in units of $c/\protect\lambda$; (b) temporal evolution of $%
\protect\gamma_x$ and $\protect\gamma$; (c) probe light reflected from the electron
layer as seen by an observer at $x/\protect\lambda=6$; (d) corresponding
spectrum. }
\end{figure}

There are two severe drawbacks, when using these flyers as REMs for
reflecting probe light. First, the flyer $\gamma(t)$ depends on time such
that the reflected pulse is chirped and has a broad spectrum. Secondly, the
mirror $\gamma_x$ is much smaller than the full $\gamma$ for $\gamma \gg 1$
(see Fig. 1(b)). In fact, we find for the Doppler shift
\begin{equation}
4\gamma _{x}^2=4\gamma^2 /(1+p_{\perp }^{2})\approx 2\gamma .  \tag{1}
\end{equation}
Apparently, the transverse momentum $p_{\perp }$, inherent to electron
motion in transverse light waves, degrades the upshift. This is true even
though the angle of electron motion relative to the laser direction, $\tan
\theta =p_{\perp }/p_{x}=2/a$, tends to vanish for large $a$.

The major result reported in this Letter is a method to overcome these two
drawbacks and to describe a practical way to generate flyers that have fixed
$\gamma $ values and $\gamma _{x}\approx \gamma $. Accordingly, they can
produce optical pulses that are Doppler-shifted by the full factor $4\gamma
^{2}$ and have a narrow spectrum. They are only limited by decreasing
reflectivity due to decay of the flyer. The way to suppress the transverse
momentum is to let the electrons interact with a counter-propagating laser
pulse. Here the central new observation is that reflection of the drive
pulse from a perfect mirror provides the ideal interaction pulse. The
corresponding configuration is sketched in Fig. 2. Due to relativistic
kinematics, the reflected pulse changes the energy $\gamma $ of the flyer
electrons only marginally, but eliminates their transverse momentum $%
p_{\perp }$ completely. While the electrons gain energy $(\Delta \gamma
)^{+}\propto a^{2}$ when co-moving with the driving pulse for a long time,
the interaction time with the reflected counter-propagating pulse is quite
short, and, accordingly, the energy loss is only of order unity, $(\Delta
\gamma )^{-}\propto 1$; this we have discussed in more detail in Ref. \cite{CTS1}.

\begin{figure}[tbp]
\resizebox{0.7\columnwidth}{!}{\includegraphics{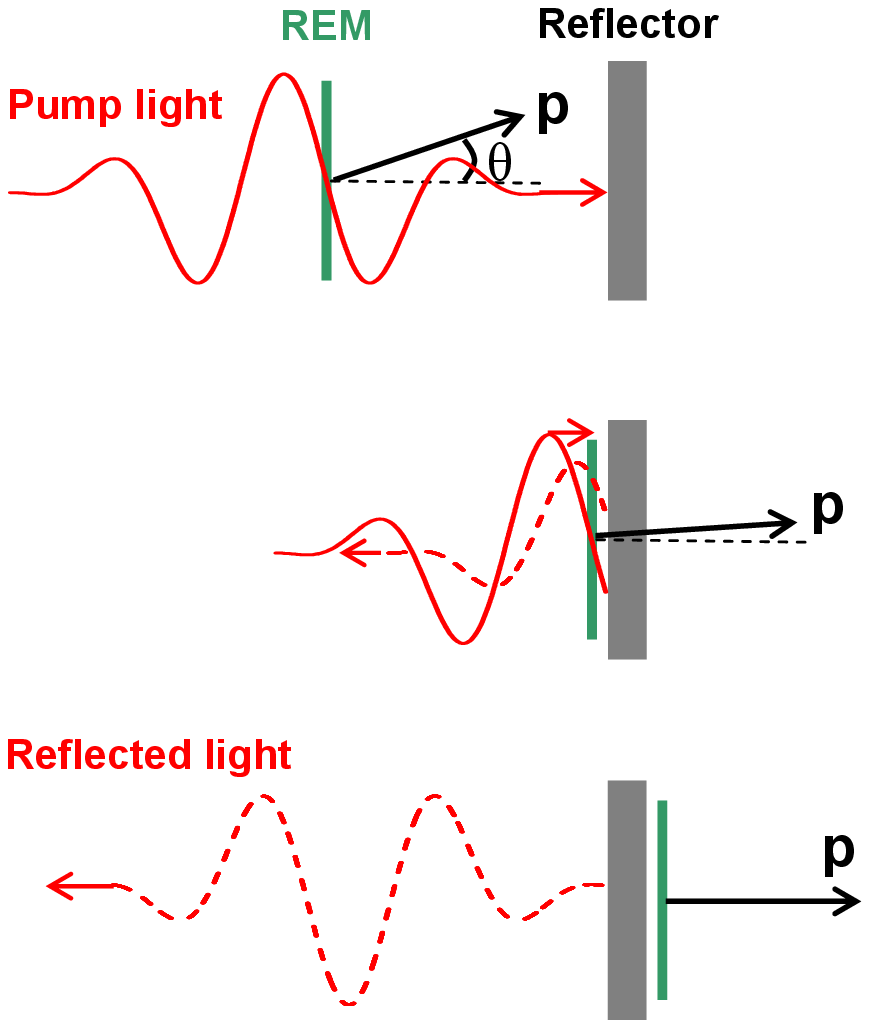}}
\caption{(color). Schematic drawing of electron layer (green) surfing on
laser drive pulse and reflector foil (grey). Electron momentum is tilted by
angle $\protect\theta$ relative to laser pulse axis. Middle part: The drive
pulse is reflected by the reflector foil and interacts a second time with
the electron layer. Lower part: The fully reflected drive pulse propagates
to the left, while the electron layer has passed the reflector foil and is
cruising to the right at constant velocity and free of transverse momentum. }
\end{figure}

Here we restrict ourselves to show explicitly only the elimination of $%
p_{\perp }$. Let us denote electric and magnetic field of the plane drive
pulse by $E^{+}=B^{+}=f(t-x)$ (fields normalized to $mc\omega _{L}/e$).
Reflection from a perfect mirror at $x=x_{R}$, taking due account of a
phase-shift $\pi $, produces the reflected pulse $%
E^{-}=-B^{-}=-f(t+x-2x_{R}) $. The electron transverse momentum obeys $%
dp_{\perp }/dt=-(E^{+}+E^{-})+\beta _{x}(B^{+}+B^{-})$. It can be obtained
as the sum $p_{\perp }=p_{\perp }^{+}+p_{\perp }^{-}$ of the momenta
resulting from the interaction with drive and reflected pulse separately.
These are described by
\begin{align}
\frac{dp_{\perp }^{+}}{dt}& =-(1-\beta _{x})f(t-x),  \tag{2a} \\
\frac{dp_{\perp }^{-}}{dt}& =(1+\beta _{x})f(t+x-2x_{R}).  \tag{2b}
\end{align}%
Using new coordinates $\tau ^{+}=t-x(t)$ with $d\tau ^{+}/dt=1-\beta _{x}$
in Eq. (2a) and $\tau ^{-}=t+x(t)-2x_{R}$ with $d\tau ^{-}/dt=1+\beta _{x}$
in Eq. (2b), one obtains the equations $dp_{\perp }^{+}/d\tau ^{+}=-f(\tau
^{+})$ and $dp_{\perp }^{-}/d\tau ^{-}=f(\tau ^{-})$, which are independent
of the particular form of the electron trajectories $x(t)$ and $\beta
_{x}=dx/dt$. Here we integrate the first one from $\tau ^{+}=0$, when the
drive pulse first touches the electron, toward $\tau ^{+}=\tau _{R}$, when
the electron hits the reflector. We also integrate the second one from $\tau
^{-}=t_{2}+x(t_{2})-2x_{R}=0$, when the reflected wave front reaches the
electron at time $t_{2}=2x_{R}-x(t_{2})$ and location $x(t_{2})$, toward $%
\tau ^{-}=\tau _{R}$. We find that the two momenta obtained from interaction
with drive pulse and reflected pulse exactly cancel each other such that the
electron emerging from behind the reflector foil has $p_{\perp }=0$ and
therefore $\gamma _{x}=\gamma $. Apparently, this result holds for each
individual electron of the flyer independent of its initial distance from
the reflector and the laser pulse shape $f(\tau )$. It is also independent
of the charge separation field which eventually causes longitudinal
dispersion of the mirror. The deeper reason for the cancelation of $p_{\perp
}$ lies in the fact that, due to planar symmetry, the invariance $p_{\perp
}=a$ still holds for the combined fields $a=a^{+}+a^{-}$ of forward-going
and reflected pulse, and therefore $p_{\perp }\rightarrow 0$ since $%
a\rightarrow 0$ when passing the reflector \cite{Huang2010}.

We have checked these results by two-dimensional particle-in-cell (2D-PIC)
simulations \cite{Wu2008}. Figure 1 exhibits electron expulsion from a thin
foil and CTS of probe light from the emerging electron flyer for the case
without reflector. Here the simulation box has a size of $10\lambda \times
20\lambda $ in $xy$ plane, and a space resolution of $1000$ cells$/\lambda $
and $800$ cells$/\lambda $ in x- and y- direction, respectively. The pump
laser pulse has a profile $a_{0}\sin ^{2}(t/T)$ with pulse duration $%
T=3\lambda /c$. The carrier-envelope-phase of the laser is set to zero, so
that the electric field maximum is at the pulse center. The transverse
profile is chosen as a super-Gaussian $\exp (-r^{4}/R^{4})$ with waist $%
R=5\lambda $. We take $a_{0}=3.5$, corresponding to an intensity of $%
I=2.6\times 10^{19}$ W/cm$^{2}$ for $\lambda =800$ nm.

Few-cycle pulses of this strength are presently becoming available \cite%
{Veisz}. The laser pulse is linearly $p$-polarized along the $y$-direction
and injected into the simulation box from the left boundary at $t=0$. For
this exploratory simulation, the initial ultrathin foil is chosen with
density $n_{e}/n_{c}=1$, thickness $L/\lambda=0.001$, and is located at $%
x_{0}=1\lambda $. The initial plasma temperature is 10eV, and the ions are
taken as immobile. The total number of macro-particles is about $5\times
10^{7}$.

The electron density evolution is shown in Fig. 1(a). An electron layer is
driven out of the foil (immobile ions at $x/\lambda =1$ not shown), and one
may notice some transverse electron motion in polarization ($y$) direction
superimposed on the drift in laser ($x$) direction. Due to the transverse
shaking, the effective area available for CTS is limited to the central zone
between $9<y/\lambda <11$. The temporal evolution of $\gamma $ and $\gamma
_{x}$ is shown in Fig. 1(b) for electrons located between $9.5<y/\lambda
<10.5$. The energy spread of the electron layer increases with time and
eventually leads to gradual longitudinal expansion. One also notices the
conspicuous difference between $\gamma $ and $\gamma _{x}$. The maximum
values $\gamma _{\max }\approx 7.0$ and $\gamma _{x,\max}\approx 1.8$ agree
with the single-electron prediction $\gamma _{x}\approx \sqrt{\gamma /2}$
(see Eq.(1)). The evolution of $\gamma (t)$ follows approximately the laser
vector potential according to the single-electron expression $\gamma
(t)=1+a(x(t),t)^{2}/2$. After $t=8\lambda /c$, the laser pulse has overtaken
the electrons, which then return to rest in agreement with the
Lawson-Woodward theorem \cite{LW1979}.

Signatures of probe light incident from the right side and backscattered
from the relativistic electron layer are also shown in Fig. 1. In the
simulation, the probe pulse is taken as a plane wave, $a_p(x,t)=a_{p0}%
\cos(k_Lx+\omega _{L}t)$, and $s$-polarized (in $z$-direction) to
distinguish it from the drive pulse. A relatively small amplitude $%
a_{p0}=0.1 $ is chosen to make sure that the scattering is linear and that
the flyer shape is not strongly perturbed. The reflected light is recorded
by a fictitious observer located at $(x,y)=(6\lambda ,10\lambda )$. The
temporal shape and spectrum of the reflected signal are shown in Fig. 1(c)
and Fig. 1(d), respectively. By adjusting the time delay, we make sure that
the fronts of both pump and probe pulse touch the production foil at the
same time. The reflected signal therefore contains the whole information on
the flyer dynamics. Since the Doppler shift $4\gamma_x(t)^2$ varies with
time, a complex oscillation pattern is observed in the reflected signal,
leading to the broad spectrum seen in Fig. 1(d); the cut-off appears at $%
\omega/\omega_L=4\gamma_x^2\approx 16$ in agreement with the peak values of $%
\gamma_x\approx 2$. The two groups of rapid oscillations represent the two $%
\gamma _{x}$ peaks in Fig. 1(b) and cause the spectral beating with period $%
\Delta\omega/\omega_0\approx 2$. Though interesting as a diagnostics of
flyer dynamics, such CTS pulses are of little value as a light source.

\begin{figure}[tbp]
\resizebox{1\columnwidth}{!}{\includegraphics{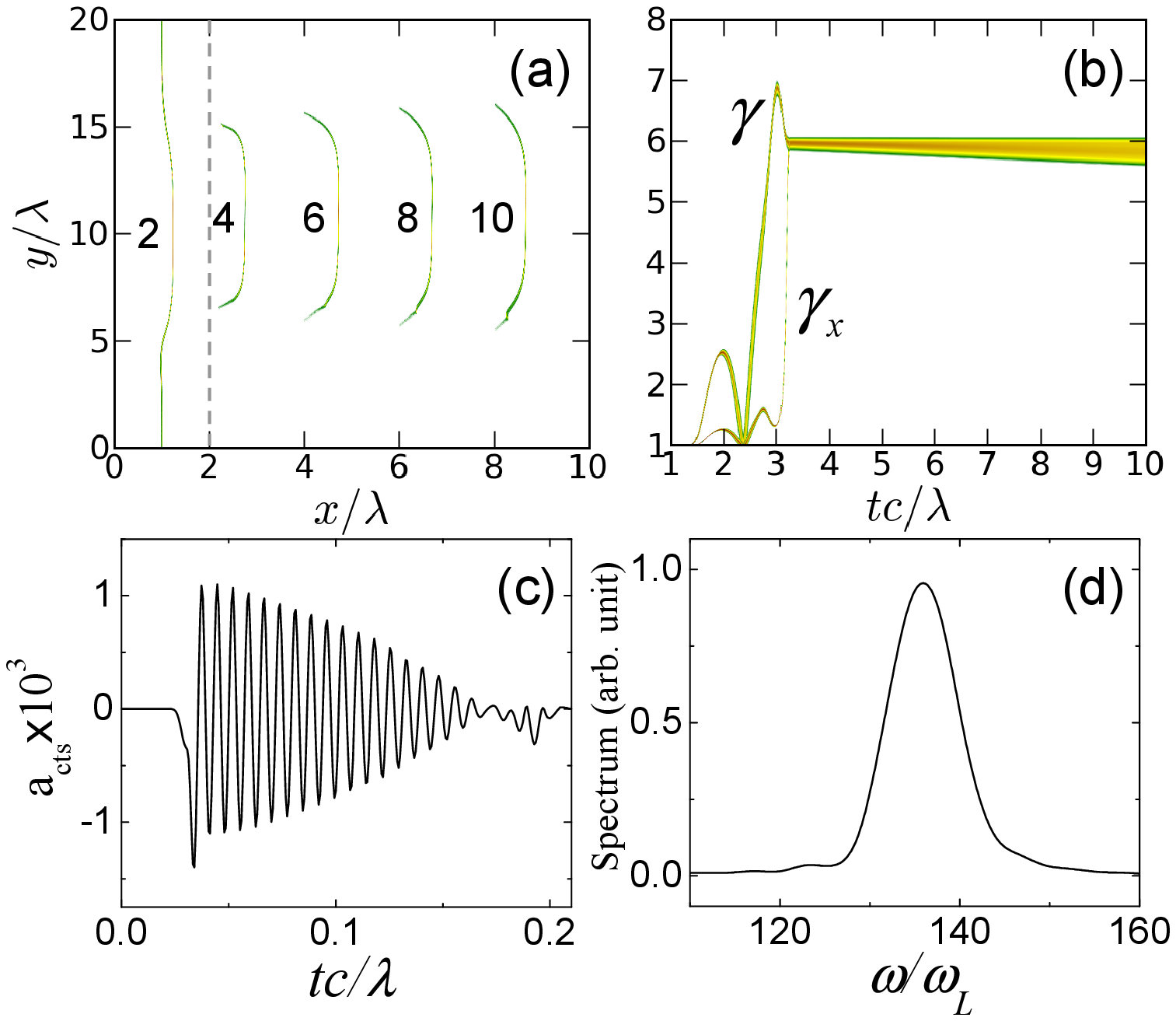}}
\caption{(color). Same as Fig. 1, but now including the 24 nm thick
reflector foil. (a) A very uniform electron layer is emerging from the
reflector foil (dashed line at $x_R/\protect\lambda=2$); (b) at about $%
t\approx 3c/\protect\lambda$, when the electron flyer passes the reflector
foil, $\protect\gamma_x$ approaches $\protect\gamma$, indicating elimination
of transverse electron momentum; the electron energy is given by nearly
constant $\protect\gamma= \protect\gamma_x \approx 5.9$, and the energy
spread increases slightly with time indicating layer expansion; (c) the
reflected probe light appears as a regular, Doppler-compressed light wave
that decays due to decreasing reflectivity of the expanding electron layer;
(d) the spectrum shows a narrow VUV pulse Doppler-shifted by $4\protect\gamma%
_x^2 \approx 139$ corresponding to a wavelength of 5.7 nm for 800 nm probe
light. For more details see text.}
\end{figure}

These results change drastically when adding the reflector foil. The
configuration is illustrated in Fig. 2. Simulation results corresponding to
those in Fig. 1, but now including the reflector, are depicted in Fig. 3.
They highlight the central results of this Letter. It is observed that the
electrons emerge from the reflector foil (marked by the dashed line in Fig.
3(a)) as a smooth dense layer in distinct contrast to the distorted shape
seen in Fig. 1. In Fig. 3(b) one sees $\gamma_x$ sharply rising and almost
exactly merging with $\gamma$ at a value of $\gamma_x= \gamma\approx 5.9$,
while $\gamma$ itself drops only by $\Delta\gamma\approx 1$. This is due to
the interaction with the reflected drive pulse. As predicted above on the
basis of single-electron motion, interaction with the reflected pulse fully
eliminates transverse electron momentum, while reducing electron energy only
marginally.

Although these sudden changes occur close to the reflector foil, it is
important to understand that they are not due to any direct action of the
reflector foil on the electrons, but are mediated indirectly by the
reflected laser field. The reflector foil is modeled here as a fully ionized
plasma slab with a density of $n_{e,R}/n_{c}=400$ and a thickness of $%
L_{R}/\lambda=0.03$. This is sufficient to fully reflect the drive pulse. A
reflectivity of $99.8\%$ is obtained in the simulation. On the other hand,
the relativistic electron layer passes the thin reflector foil practically
unperturbed. The relative energy loss due to Coulomb collisions is found to
be negligible in the order of $10^{-5}$.

It appears that a uniform relativistic electron flyer with constant $\gamma$
is produced that can now act as a CTS mirror. Backscattering of probe light
by this mirror has been simulated, using the same parameters as for the
results of Fig. 1, but now increasing space resolution in $x$-direction to
1500 cells/$\lambda$ in order to resolve the CTS signal of much higher
frequency and also enlarging the box length in $x$-direction to $15\lambda$.
The reflected signal now consists of a regular wave train strongly
compressed by the moving mirror (see Fig. 3(c)) and with a spectral peak
(see Fig. 3(d)), Doppler-shifted by the factor $4\gamma_x^2=4\gamma^2\approx
139$ in best agreement with $\gamma=5.9$. Here the observation point has
been moved to $(x,y)=(13\lambda ,10\lambda )$ to record the full CTS signal.
Although the probe light is incident continuously, the CTS signal decays due
to the expansion of the flyer clearly seen in Fig. 3(c). For an incident
wavelength of $\lambda=800$ nm, the X-ray pulse has a duration of about 500
attoseconds and a central wavelength of 5.7 nm. The maximum amplitude of $%
a_{cts}=1.4\times 10^{-3}$ corresponds to the intensity of $4.2\times
10^{12} $ W/cm$^{2}$ and to a reflectivity of about $10^{-6}$. By slightly
changing the parameters, we expect that a pulses of $10^9$ coherent X-ray
photons are possible covering the range of the water window.

The coherent reflectivity of these electron flyers has been studied in Ref.
\cite{CTS2}. For a flyer density distribution $n_{e}(x)=n_{e,0}S(x/L_{c})$,
given by a symmetric shape function $S(x/L_{c})=S(-x/L_{c})$ with
characteristic length $L_{c}$, the CTS amplitude in normal direction can be
obtained as
\begin{equation}
a_{cts}=a_{p0}\frac{n_{e,0}}{2\gamma n_{c}}F(\xi ).  \tag{3}
\end{equation}%
where $F(\xi )$ is the Fourier transform
\begin{equation}
F(\xi )=\int_{-\infty }^{\infty }S(\chi /\xi )\cos (2\chi )d\chi .  \tag{4}
\end{equation}%
Here $\chi \equiv k_{L}^{\prime }x^{\prime }$ and $\xi \equiv k_{L}^{\prime
}L_{c}^{\prime }$ denote the normalized x coordinate and layer thickness in
the rest frame of the electron flyer; in the lab frame $\xi $ is given by $%
\xi =\gamma ^{2}(1+\beta )k_{L}L_{c}$. A similar form-factor $F(\xi )$
appears in the theory of coherent synchrotron radiation \cite{Williams1989}.
For a Gaussian profile $S(x)=\exp (-x^{2}/L_{c}^{2})$, we find $F(\xi )=%
\sqrt{\pi }\xi \exp (-\xi ^{2})$. Apparently, for such a shape the reflected
signal vanishes exponentially as soon as $\xi \gg 1$, i.e. when the flyer
becomes thicker than the wavelength of the reflected, $L_c\gg \lambda /4\gamma
^{2}$. For the parameters studied above, we have $n_{e,0}=0.252n_{c}$, $%
L_{c}=0.0017\lambda $ (Gaussian fitting), $\gamma =5.96$, and obtain $%
a_{cts}=$ $1.6\times 10^{-3}$ in reasonable agreement with the PIC result
shown in Fig. 3(c). The conversion ratio of pulse energy is found from Eq.
(3) as
\begin{equation}
\alpha =\frac{n_{e,0}^{2}}{4n_{c}^{2}}\frac{F(\xi )^{2}}{\gamma ^{4}(1+\beta
)^{2}}.  \tag{5}
\end{equation}

In conclusion, we have described a new method to make high-quality,
micro-scale, relativistic electron mirrors. Dense electron layers, driven
out from nanometer-thick production foils by few-cycle laser pulses, are
significantly improved by introducing an additional reflector foil that
reflects the drive laser pulse, but lets the electrons pass unperturbed. As
a result, very uniform relativistic electron layers are obtained, freely
cruising at fixed high-$\gamma$ values precisely in drive-laser direction
and with zero transverse momentum. Here we have discussed coherent Thomson
backscattering from these layers to generate monochromatic, coherent, soft
x-ray pulses, Doppler-shifted by a factor $4\gamma^2$. The goal was to
explain basic features in terms of single-electron dynamics and verifying
the scheme by 2D-PIC simulation. The next step will be experimental
demonstration. Though demanding, both the generation of the required
high-contrast laser pulses and the fabrication of the micro-scale double
foil targets should be within range in the next future. Successful
demonstration would provide a new versatile tool for generating powerful
laser-like X-ray pulses on a micro-scale.

\begin{acknowledgments}
The authors are grateful to Dr. Lin Yin for useful discussions and comments
and Dr. Chengkun Huang for a very helpful comment. J. Meyer-ter-Vehn
acknowledges support by the Munich center for Advanced Photonics (MAP) and
by the Association EURATOM - Max-Planck-Institute for Plasma Physics.
\end{acknowledgments}

\end{document}